\documentclass[10pt]{revtex4}
\usepackage{graphics}
\usepackage{amsfonts}
\usepackage{amssymb}
\usepackage{epsfig}
\usepackage{amsmath}

\begin{document}

\title{Entanglement and  statistics in Hong-Ou-Mandel interferometry}
\author{Vittorio Giovannetti}
\affiliation {NEST CNR-INFM \& Scuola Normale Superiore,
Piazza dei Cavalieri 7, I-56126 Pisa, Italy
\\ \\
{\em e-mail address}: v.giovannetti@sns.it} 
\date{\today}
\begin{abstract}
Hong-Ou-Mandel interferometry allows one to detect the presence
of entanglement in two-photon input states. 
The same result holds for two-particles input  states which
obey to Fermionic statistics. In the latter case however  
anti-bouncing introduces qualitative differences in the
interferometer response. 
This effect is analyzed in a {\em Gedankenexperiment}  where
the particles entering the interferometer are assumed to 
belong to a one-parameter family of 
{\em quons} which continuously interpolate between the
 Bosonic and Fermionic statistics.
\end{abstract}
\maketitle

{\bf Topic:} Quantum measurements, Quantum optics and related technologies,
Quantum entanglement and non-locality,
New perspectives for Foundations of Quantum Mechanics from Quantum Information
%\newpage
\section{Introduction}

Interferometry is a widely used
quantum optical technique for characterizing the non classical behaviors 
of light, like squeezing or entanglement (see for instance~\cite{WALLS}
and references therein).
In particular the presence of entanglement in states of two
multi-mode photons
can be detected by means of 
Hong-Ou-Mandel (HOM) interferometry~\cite{HOM,ERDMANN}.
Indeed, the visibility of the coincidence  counts at the 
output ports  of such interferometer allow to detect entanglement in
 the incoming two-photon states: 
the visibility of separable two-photon inputs is in fact
upper bounded
by half of the maximum achievable visibility 
(e.g. the visibility associated with the entangled {\em twin-beam state} 
emerging from a monochromatic pumped $\chi^{(2)}$ non-linear crystal).
In the context  of mesoscopic physics~\cite{BEN} 
a Fermionic analog of the HOM interferometer has been
recently proposed in Ref.~\cite{NOSTRO}: here 
metallic leads play the role of photonic multi-mode optical paths 
and current-current correlations play the role of the 
coincidence counts. 
Notwithstanding the change in the input particle statistics
(from Bosonic to Fermionic) and the corresponding 
transition from
an underlying bouncing behavior to an underlying 
anti-bouncing behavior, 
it turns out that  the HOM 
interferometer retains its full entanglement
witnessing capability. 
As a matter of fact the output signals of the Fermionic 
HOM implementation~\cite{NOSTRO} 
can be easily mapped  into the output signals of the standard
Bosonic implementation: the
only difference being a different definition of the 
``regions'' which are accessible or not accessible to
 separable inputs.
Motivated by the recent interest in the relations 
between entanglement,  statistics and interferometry~\cite{BOSE,NOSTRO,LIM,BEN,LAVAGNO,ZHANG,HBT,BURKARD}
we present a  simple theoretical model which
permits  to analyze them 
in a broader theoretical context. In particular, we discuss an hypothetical
HOM set-up where the two particles entering the interferometer 
obey to generalized quantum statistics~\cite{GREEN,GREEN1}. This 
allow us 
to interpolate continuously
between the Bosonic HOM interferometer and its 
Fermionic counterpart.

We start by briefly reviewing the standard Bosonic HOM interferometer.
Then we introduce the {\em quon} algebra~\cite{GREEN,GREEN1} 
and discuss a HOM-{\em Gedankenexperiment} which operates on particles 
obeying to the corresponding deformed statistics.

\section{Using Hong-Ou-Mandel interferometer to detect entanglement in
multi-mode two-photon states}\label{s:quons}

\begin{figure}[t]
\centerline{\psfig{file=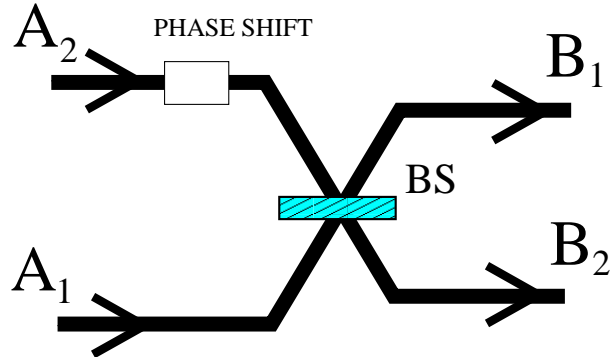,width=8 cm}}
\vspace*{8pt}
\caption{Schematic of the Hong-Ou-Mandel interferometer:
the two-particle state $|\Psi\rangle$ of Eq.~(\ref{input}) 
 enters the set-up from the multi-mode 
input ports $A_1$ and 
$A_2$ and combines at the 50/50 beam-splitter BS. 
Coincidence counts~(\ref{coinc}) are measured at  the multi-mode 
output ports $B_1$ and $B_2$
for different values of the controllable phase-shift delay introduced 
in white-box area.}
\label{figu1}
\end{figure}

The  prototypical  HOM set-up~\cite{HOM} is sketched in
Fig.~\ref{figu1}.
Here two multi-mode photons 
enter the interferometer following, respectively, the 
optical paths associated with the input  ports $A_1$ and $A_2$.  
Their (possibly entangled) state is described by the vector
\begin{eqnarray}
|\Psi\rangle =\sum_{k_1,k_2} \Phi(k_1,k_2) a_1^\dag(k_1)a_2^\dag(k_2) |
\O \rangle
\label{input}\;,
\end{eqnarray} 
where $|\O\rangle$ is the vacuum and where
$a_j(k)$ is the photonic annihilation operator
associated with $k$-th mode of the port $A_j$ which
obeys standard 
Bosonic commutation relations, i.e.
\begin{eqnarray}
\left[a_j(k), a_{j^\prime}(k^\prime) \right] &=& 0 \label{prima} \\
\left[ a_j(k), a_{j^\prime}^\dag(k^\prime) \right]  
&=& \delta_{jj^\prime} \; \delta_{kk^\prime} \label{seconda}\;.
\end{eqnarray}
In  these expressions $k$ labels
the  different frequencies   
of the modes propagating along 
the optical paths $A_{1,2}$ (for the sake of simplicity
the   entering  signals are supposed to have definite
polarization)
and $\Phi(k_1,k_2)$ is the two-photon spectral 
amplitude 
which satisfies the normalization condition
\begin{eqnarray}
\sum_{k1,k_2} |\Phi(k_1,k_2)|^2 =1 \;. \label{norma}
\end{eqnarray}
Within the interferometer
the modes of port $A_2$ undergo to phase-shifts transformations
introduced through 
controllable delays (represented by the
white box of Fig.~\ref{figu1}) which  transform $a_2(k)$
into $a_2(k) e^{i\varphi_k}$.  The two optical paths then
interfere at the 50/50 beam-splitter BS 
where photo-coincidence $C_{12}$ are measured
at the output ports $B_1$ and $B_2$.
We will see that 
the presence of entanglement 
in the photon input state $|\Psi\rangle$ can
be revealed by studying $C_{12}$ 
as a function of the controllable delays.

%%%%%%%%%%%%%%%%%%%%%%%%%%%%%%%%%%%%%%%%%%%%%%%%%%%%%%%%%%%%%%%%%%%%
%%%%%%%%%%%%%%%%%%%%%%%%%%%%%%%%%%%%%%%%%%%%%%%%%%%%%%%%%%%%%%%%%%%%%
Assuming perfect detection efficiency and considering 
the limit of long detection times $T$
the average coincidence counts of the photo-detectors 
operating on $B_1$ and $B_2$ can be expressed as
\begin{eqnarray}
C_{12} = \lim_{T\rightarrow \infty} \int_{0}^T  d t_1
\int_{0}^T d t_2 \; \frac{
\langle \Psi | I_{B_1}(t_1)  I_{B_2}(t_2) 
|\Psi \rangle}{T^2} \;, \label{coinc} 
\end{eqnarray}
where $I_{B_{1,2}}(t)$ are the normalized output intensity operators
defined by  
\begin{eqnarray}
I_{B_j}(t) \equiv \sum_{k_1} \sum_{k_2}
e^{i (\omega_{k_1}-\omega_{k_2})t} 
\label{photonumber} b_j^\dag(k_1)  b_j(k_2) 
\;,
\end{eqnarray}
with $b_j(k)$
being  the Bosonic annihilation operator associated with the 
$k$-th outgoing mode of port
$B_j$. 
Evaluating the integrals and the limit of Eq.~(\ref{coinc}) gives
\begin{eqnarray}
C_{12} = \langle \Psi | \;N_1 N_2 \;|\Psi \rangle  \;, \label{coinc1} 
\end{eqnarray}
where for $j=1,2$, $N_j$ is the total photon number operator 
of the $B_j$ output port, i.e.
\begin{eqnarray}
N_j = \sum_k n_j(k)  \;, \label{tot}
\end{eqnarray}
with $n_j(k)= b_j^\dag (k) b_j(k)$ being the number operator
of the $k$-th mode.
Analogously the average output photo-counts at $B_1$ and $B_2$ yield
\begin{eqnarray}
i_{j} \equiv   \lim_{T\rightarrow \infty} \int_{0}^T d t \; \frac{
\langle \Psi | I_{B_i}(t) 
|\Psi \rangle}{T} = 
\langle \Psi | \; N_j \;
|\Psi \rangle  \;. \label{corrente} 
\end{eqnarray}
The operators $b_j(k)$ and $a_j(k)$ are connected through the
input-output relations of the HOM interferometer, i.e.
\begin{eqnarray}
b_1(k) &=& \left[ a_1(k) + e^{i\varphi_k} a_2(k) \right] /\sqrt{2} 
\nonumber 
\\
b_2(k) &=& \left[ a_1(k) - e^{i\varphi_k} a_2(k) \right] /\sqrt{2}\;,
\label{inout}
\end{eqnarray}
with $\varphi_k$ being the controllable phase-shift 
and with the coefficient $1/\sqrt{2}$ 
 coming from the 50/50 beam-splitter
transformation. Notice that Eq.~(\ref{inout}) couples only those 
modes of $A_1$ and $A_2$ which share the same value of 
$k$ and that 
$\varphi_k$  depends explicitly upon such index.
[In Refs.~\cite{HOM,ERDMANN} for instance
it is
$\varphi_k = \omega_k \tau$
with $\omega_k$ being the frequency 
of the $k$-th mode and with $\tau$ being a tunable delay.]
With the help of the above definitions
one can show that the average photon number~(\ref{corrente}) 
at  each  of the 
two output ports equals $1$. On the other hand Eq.~(\ref{coinc1}) 
yields
\begin{eqnarray}
C_{12} = \frac{1 - w}{2}\;, \label{BOSONI}
\end{eqnarray} 
with 
\begin{eqnarray}
w =  
 \sum_{k_1,k_2} \Phi^*(k_1,k_2) \Phi(k_2,k_1) 
e^{i (\varphi_{k_1} - \varphi_{k_2})}\;\label{theg} \;.
\end{eqnarray}
It is possible to verify that
for a suitable choice of the delays
 $\varphi_k$ and the
spectral function $\Phi(k_1,k_2)$, the real quantity
$w$ can take any values in the interval $[-1,1]$. 
Correspondingly $C_{12}$
can take values over the interval $[0,1]$.
In the 
twin-beam state case of Ref.~\cite{HOM,ERDMANN}, for instance,  
the plot of $w$  with respect to  the controllable delay $\tau$
shows the so called {\em Mandel dip}, i.e. it nullifies for $\tau=0$ and
approaches $1$ for sufficiently large $\tau$. This implies that,
by defining the 
visibility $V$ of $C_{12}$ 
as the difference between the maximum and minimum  values
it assumes when varying $\varphi_k$, 
the maximum achievable value of $V$ for the two-particles
state $|\Psi\rangle$ is $1$.

\begin{figure}[t]
\centerline{\psfig{file=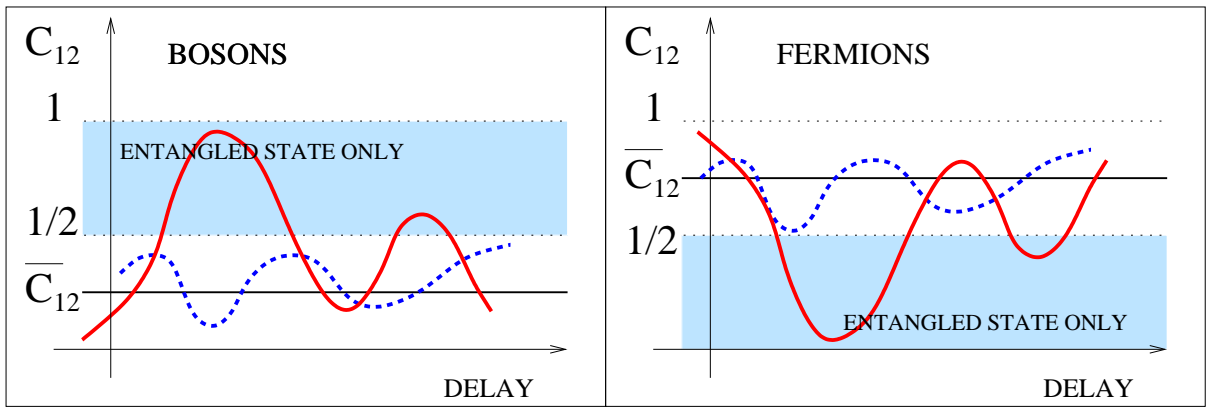,width=16 cm}}
\vspace*{8pt}
\caption{Pictorial representation of the coincidence counts~(\ref{coinc1}) dependence upon the interferometric delay $\varphi_k$ in the
HOM interferometer of Fig.~\ref{figu1}. Left: Bosonic case.
Among the two-particle states $|\Psi\rangle$ of Eq.~(\ref{input})
only the entangled one (represented by the 
continuous curve) can have $C_{12}>1/2$;
separable states (dotted curve) 
stay always below the $1/2$ threshold: the gray area above $1/2$ is
not accessible to them.   
Right: Fermionic case. Here the  area which is not accessible to
separable states is the region below $1/2$ (see Ref.~\cite{NOSTRO}
for details). The continuous lines represent the average
value~(\ref{cmedio}) of $C_{12}$.}
\label{figu2}
\end{figure}

On the other hand, consider what happens when $|\Psi\rangle$ factorizes with respect to $A_1$ and $A_2$. In this case the spectral amplitude
has the form $\Phi(k_1,k_2) = \phi_1(k_1) \phi_2(k_2)$ and Eq.~(\ref{theg})
becomes
\begin{eqnarray}
w = w_{\mbox{sep}}\equiv | \sum_{k} \phi_1^*(k) \phi_2(k) 
e^{i\varphi_k}|^2 \geqslant 0 \;.\label{thegsep}
\end{eqnarray}
Consequently, the coincidence counts $C_{12}$ of 
 factorizable input states $|\Psi\rangle$ 
are limited into the interval $[0,1/2]$ and their maximum allowed
visibility (i.e. $1/2$) is only one half of the
maximum achievable value (see Fig.~\ref{figu2}).
As discussed in Ref.~\cite{NOSTRO} the above result generalize to
any (non necessarily pure) two-photon input density matrices 
which are separable with respect
to $A_1$ and $A_2$. Therefore we can use $C_{12}$  as
an entanglement witness for two-photon input states. Indeed it is 
sufficient to repeat the coincidence measurements 
for different values of the controllable delay:
by realizing a value of $C_{12}$ or 
a  visibility $V$ which are strictly greater
than $1/2$ one can conclude that the 
input state of the system was 
 entangled.

%%%%%%%%%%%%%%%%%%%%%%%%%%%%%%%%%%%%%%%%%%%%%%%%%%%%%%%%%%%%%%%%%%%%
%%%%%%%%%%%%%%%%%%%%%%%%%%%%%%%%%%%%%%%%%%%%%%%%%%%%%%%%%%%%%%%%%%%%%
\subsection{Bosons vs. Fermions}
In Ref.~\cite{NOSTRO} a Fermionic
implementation of the HOM interferometer  has
been proposed. There, effective two-electron
states analog to the two-photon input states of Eq.~(\ref{input}) 
originate from a solid-state entangler~\cite{BEN} and propagate along
metallic leads which take the place of the optical ports 
$A_1$, $A_2$, $B_1$ and $B_2$ of Fig.~\ref{figu1}. Moreover,
the coincidence counts $C_{12}$ of Eq.~(\ref{coinc}) 
is replaced by the current-current correlator of the outgoing
leads $B_1$ and $B_2$.  
We refer the reader to Ref.~\cite{NOSTRO} for a 
detailed discussion of the physical assumptions 
underlying such a scheme.
For the purposes of the present manuscript it is sufficient to
observe that  
an effective but rigorous description of such interferometer
is obtained from the Eqs. (\ref{input}), (\ref{coinc1}) and 
(\ref{corrente})
by simply replacing the
Bosonic commutation rules~(\ref{prima}) and (\ref{seconda})
with their Fermionic counterpart, i.e. 
\begin{eqnarray}
\left\{ a_j(k), a_{j^\prime}(k^\prime) \right\} &=& 0 \label{primaf} \\
\left\{ a_j(k), a_{j^\prime}^\dag(k^\prime) \right\}  
&=& \delta_{jj^\prime} \; \delta_{kk^\prime} \label{secondaf}\;,
\end{eqnarray}
with $\{r,s\} = rs + sr$ being the anti-commutator of the operators
$r$ and $s$. 
With this substitution
the average output electron-numbers  $i_{1,2}$ of Eq.~(\ref{corrente})
remain equal to $1$. However the expression~(\ref{BOSONI}) for the 
coincidence-counts $C_{12}$ is replaced by
\begin{eqnarray}
C_{12} = \frac{1 + w}{2} \label{FERMIONI}
\end{eqnarray} 
with $w$ as in Eq.~(\ref{theg}). 
Since the dependence of $w$ upon the factorizable properties of 
the input states $|\Psi\rangle$
is the same as in the Bosonic case, one can still use
the visibility $V$ as a signature of entanglement.
Indeed also in the Fermionic case, 
entangled inputs can have visibility $V$  greater than $1/2$, while
separable have always visibility $V$ smaller than or equal to $1/2$.
In the case where the label $k$ refers to the spin degree of freedom of the
incoming electron  this effect was  first noted in  Ref.~\cite{BURKARD}.

The sign
difference between Eqs.~(\ref{BOSONI}) and~(\ref{FERMIONI})
implies that separable states of Fermions  are
forced to have $C_{12}$ greater than or equal to $1/2$ while
separable states of Bosons have $C_{12}$ smaller than or equal to $1/2$
(see Fig.~\ref{figu2}). This phenomenon
can be interpreted in terms of the different bouncing  
and anti-bouncing attitudes of Bose and Fermi statistics. 
Indeed due to the
exclusion principle one expects the
coincidence counts $C_{12}$ of Fermions to be typically
higher than the corresponding Bosonic coincidence counts.
To make this a quantitative statement let us consider the average
value of $C_{12}$ over all possible two-particle states of Eq.~(\ref{input}), i.e. over all the normalized two-particles
spectral amplitudes $\Phi(k_1,k_2)$.
A simple symmetric argument can be used to
show
\begin{eqnarray}
\overline{w} =  
\int \; d \mu(\Phi)\; \left[ 
\sum_{k_1, k_2} \Phi(k_1,k_2) 
\Phi^*(k_2,k_1) e^{i( \varphi_{k_1}-
\varphi_{k_2})} \right] = 1/M \label{thegaverage}\;,
\end{eqnarray}
with $M$ being the number of the $k$-th modes.
Therefore the average coincidence counts is
\begin{eqnarray}
\overline{C_{12}} = 
\left\{ \begin{array}{lll}
1/2 - 1/(2M)  & & \mbox{Bosons} 
\\
\\1/2 + 1/(2M) && \mbox{Fermions.}
\end{array}
\right. \label{cmedio}
\end{eqnarray}
It is worth noticing that entangled input states are, 
to some extent, insensitive to the complementarity behavior of
Eq.~(\ref{cmedio}): indeed, as shown in Fig.~\ref{figu2}, 
there are no regions which are strictly forbidden to them by the
particle statistics.

\section{HOM interferometry with quons}\label{quons}

In this section a {\em Gedankenexperiment} is analyzed where 
the particles entering the HOM interferometer are assumed  
to be quons~\cite{GREEN,GREEN1}. 
These have been introduced by Greenberg as an example of 
continuous interpolation between
Bose and Fermi algebras (see Refs.~\cite{CHAI,LAVAGNO}
for alternative definitions). 
We will use them to describe the
sharp transition from the left part to 
the right part of Fig.~\ref{figu2} 
in terms of a continuous deformation of the particle statistics.

For $q\in [-1,1]$ the quon algebra is 
obtained by 
replacing the 
relations~(\ref{seconda}) and~(\ref{secondaf})  with the identity
\begin{eqnarray}
a_j(k) a_{j^\prime}^\dag(k^\prime) - q \;
a_{j^\prime}^\dag(k^\prime) a_j(k)
&=& \delta_{jj^\prime} \; \delta_{kk^\prime} \label{quon}\;,
\end{eqnarray}
and by neglecting~\cite{NOTA} the relations~(\ref{prima}) and~(\ref{primaf}).
For $|q|<1$, Eq.~(\ref{quon})
can be used to define a valid
non-relativistic field theory
but poses serious problems  in
deriving a reasonable relativistic quantum field theory ~\cite{GREEN,GREEN1}.
A proper Fock-like representation can be derived
upon enforcing the vacuum condition 
\begin{eqnarray}
a_j(k) |\O\rangle = 0\;,
\label{vacuum}
\end{eqnarray}
for all $j$ and $k$~\cite{GREEN,GREEN1}.
In particular, it can be shown that for $|q|<1$
the squared norms of all vectors made by polynomials of $a_j^\dag(k)$
acting on the vacuum state $|\O\rangle$ are strictly positive.
For $q=1$ and $q=-1$ instead the squared norms are never negative
and nullify for those configurations which are, respectively, 
totally symmetric and totally antisymmetric under permutations.
This allows us to recover the Bosonic and Fermionic
statistics  as limiting cases of Eq.~(\ref{quon})  without explicitly 
imposing the conditions (\ref{prima}) and (\ref{primaf}), respectively.
Of particular interest is also the $q=0$ case whose
statistics can be interpret~\cite{GREEN} as quantum version
of the Boltzmann statistics, based on a system
of identical particles having 
infinite number of internal degree of 
freedom.

The possibility of defining a proper Fock-like structure for the quon
algebra~(\ref{quon}) implies the existence of number operators 
$n_j(k)$ which satisfy the standard commutation relations
\begin{eqnarray}
\left[ n_j(k), a_{j^{\prime}}(k^\prime) \right] = -\delta_{jj^\prime}
\; \delta_{kk^\prime} \; a_j(k)
\label{number}\;,
\end{eqnarray}  
and which reduce
to $a_j^\dag(k) a_j(k)$ in the  
Bosonic and Fermionic limits~\cite{GREEN,GREEN1}.
 %%%%%%%%%%%%%%%%%%%%%%%%%%%%%%%%%%%%%%%%%%%%%%%%%%%%%%%%%%%%%%%%%%%%
\begin{figure}[t]
\centerline{\psfig{file=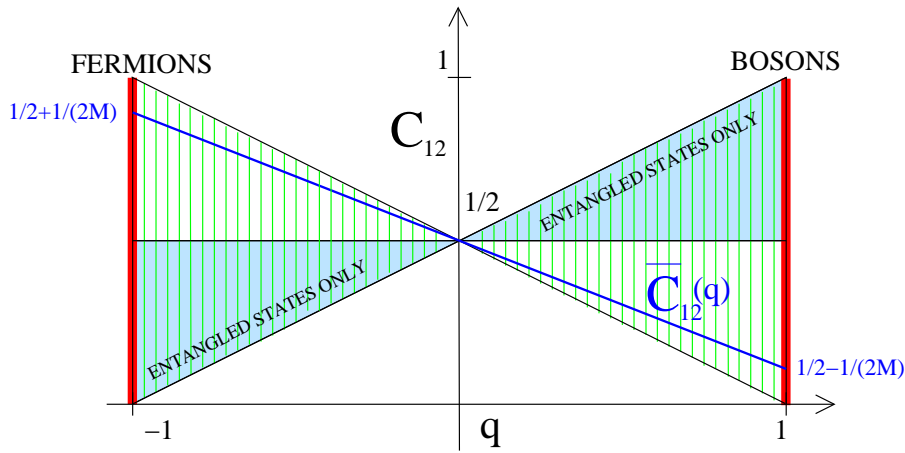,width=12 cm}}
\vspace*{8pt}
\caption{Plot of the maximum and minimum values of $C_{12}$ of Eq.~(\ref{coincq1}) allowed
for a two-particle quon state as a function of the parameter 
$q$ of the quon algebra. The shaded region represents the
allowed values~(\ref{allowed}) of $C_{12}$. The gray region 
is accessible only to non factorizable
input configurations -- see Eq.~(\ref{allowed1}). The continuous
line represents the average value~(\ref{cmedioq}) 
of $C_{12}$  over
all possible input two-particles states $|\Psi\rangle$.
For $q=1$ and $q=-1$ the boundaries 
coincides with those of  Bosonic
and Fermionic case respectively. 
The Boltzmann statistics~\cite{GREEN} $q=0$ admits only the value  
$1/2$.}
\label{figu3}
\end{figure}
%%%%%%%%%%%%%%%%%%%%%%%%%%%%%%%%%%%%%%%%%%%%%%%%%%%%%%%%%%%%%%%%%%%%%
In the following we will adopt a pragmatic point of view
assuming 
that the equations of the previous section
yield a legitimate description of our quon HOM interferometer.
This is in part justified by the fact  that the input-output 
relations~(\ref{inout}) map the quon annihilation operators
$a_j(k)$ into annihilation operators $b_j(k)$ which still
satisfy the quon algebra~(\ref{quon}) and the vacuum condition~(\ref{vacuum}). The only technical problem of our approach 
comes from the fact that
for $|q|<1$ Eq.~(\ref{coinc1}) is not necessarily a real quantity
(it is a consequence of the fact that no commutation relation
between $a_1(k_1)$ and $a_2(k_2)$ is defined~\cite{NOTA}).
Consequently we replace the product  $N_1 N_2$
with its Hermitian part, i.e.
\begin{eqnarray}
C_{12} \equiv  \langle \Psi | \frac{ N_1 N_2 + N_2 N_1}{2} | \Psi\rangle =
 \Re{e}  \; \langle \Psi |  N_1 N_2 | \Psi\rangle \;. \label{coincq}
\end{eqnarray} 
[In this expression $N_{1,2}$ is defined through Eq.~(\ref{tot})
in terms of the number
operators $n_j(k)$ of~Eq.~(\ref{number})].
The right hand side term of Eq.~(\ref{coincq}) 
can now be computed by inverting the
transformation~(\ref{inout}) and expressing the input state
$|\Psi\rangle$ of Eq.~(\ref{input}) in terms of the
operators $b_{j}(k)$.  
With the help of the relations (\ref{number}) and (\ref{quon})
and using the vacuum condition~(\ref{vacuum}) one can then verify
that the quon ``coincidence counts''  of the $q$-algebra are~\cite{NOTA10},
 \begin{eqnarray}
C_{12}(q) =\frac{1- q \;  w}{2} \label{coincq1}
\end{eqnarray}
with $w$ as in Eq.~(\ref{theg}). Analogously one can verify that
for normalized input states Eq.~(\ref{norma}) still
holds and that 
the average number $i_{1,2}$ is equal to $1$ for all $q$.
As expected, for $q=\pm 1$ Eq.~(\ref{coincq1}) reduces to the Bosonic and
Fermionic case. More interestingly
 for $q=0$, $C_{12}(q)$  does not depend upon
the two-particle input state 
$|\Psi\rangle$ and has constant value $1/2$. 
This is exactly what 
one would expect from a classical model
of the interferometer confirming the Boltzmann interpretation~\cite{GREEN} of the $q=0$ algebra.

Taking into account that for generic input state 
one has $w\in [-1,1]$ the following bounds can be derived:
\begin{eqnarray}
\begin{array}{cll}
\frac{1-q}{2} \leqslant C_{12}(q) \leqslant \frac{1+q}{2}  &
&\mbox{for $q\in [0,1]$} \\ \\
\frac{1+q}{2} \leqslant C_{12}(q) \leqslant \frac{1-q}{2}  & &
\mbox{for $q\in [-1,0]$} \;. 
\end{array} \label{allowed}
\end{eqnarray}
On the other hand for factorizable amplitudes
 $\Phi(k_1,k_2)$, 
the function $w$ obeys Eq.~(\ref{thegsep}). Therefore for these states
one has
\begin{eqnarray}
\begin{array}{cll}
\frac{1-q}{2} \leqslant C_{12}(q) \leqslant \frac{1}{2}  &
&\mbox{for $q\in [0,1]$} \\ \\
\frac{1}{2} \leqslant C_{12}(q) \leqslant \frac{1-q}{2}  & &
\mbox{for $q\in [-1,0]$} \;.
\end{array}\label{allowed1}
\end{eqnarray}
Finally, we can use Eq.~(\ref{thegaverage}) 
to compute the average value of
value of $C_{12}$ with respect all possible two-particle input
states $|\Psi\rangle$, i.e.
\begin{eqnarray}
\overline{C}_{12}(q) = 1/2 -q/(2M) \label{cmedioq}\;.
\end{eqnarray}
The above constraints and Eq.~(\ref{cmedioq})
have been plotted in Fig.~\ref{figu3}:
they indicate that the transition from the Bosonic regime to the
Fermionic regime is characterized by a critical
point at $q=0$. Here the values $C_{12}> 1/2$  which for $q>0$
were pertinent to non factorizable input states $|\Psi\rangle$,
become accessible to factorizable configurations. 
At the same time,
however the values  $C_{12}<1/2$ become inaccessible to them.
Following the discussion of the previous section this effect
can be interpreted as a continuous transition from  a bouncing
behavior to an anti-bouncing behavior with $q=0$ corresponding to
the classical ``neutral''  point.
Moreover, Eq.~(\ref{cmedioq}) provides an indirect confirmation 
of the interpretation~\cite{GREEN} 
of the $q=0$ case in terms of the statistics a system
of identical particles having infinite number of internal
degree of freedom. In effect according to such expression 
the limit $M\rightarrow \infty$ and $q\rightarrow 0$ of
the average coincidence counts $\overline{C}_{12}(q)$ are
 equivalent.

\section{Conclusions}
We analyzed 
how different quantum statistics affect 
HOM interferometry. In particular we focused on the 
interferometer response to 
initially separable/entangled inputs.
By introducing a family of $q$-deformed algebras
we interpolated between the Bosonic and Fermionic regimes
which were previously discussed in Ref.~\cite{NOSTRO}.
Our results indicate a progressive attenuation of the
bouncing behavior 
when  moving from the Bose statistics
to the Fermi statistics 
a progressive attenuation of the
bouncing behavior. 

\acknowledgments We thank Prof. Rosario Fazio and Dr. Diego
Frustaglia 
for comments and criticisms. This work
 was in part supported by the
Quantum Information research program of Centro di Ricerca
Matematica Ennio De Giorgi of Scuola Normale Superiore.


\begin{thebibliography}{99}
\bibitem{NOTA}For $q\neq \pm 1$, no (deformed) 
commutation rules can be imposed
among  $a_j(k)$ and $a_{j^\prime}^\dag(k^\prime)$ (see 
Ref.~\cite{GREEN} for details).
\bibitem{NOTA10}It should be stressed that to derive
 Eq.~(\ref{coincq1}) the exact expression
of the $n_j(k)$ in term of the annihilation operators is not
required. Only the
commutation relation~(\ref{number}) is needed.
\end{thebibliography}
\end{document}